# Discovering the Sky at the Longest wavelengths with a lunar orbit array

Xuelei Chen[1*], Jingye Yan[2], Li Deng[2], Fengquan Wu[1], Lin Wu[2], Yidong Xu[1], Li Zhou[2]

[1]*National Astronomical Observatories, Chinese Academy of Sciences, 20A Datun Road, Beijing 100101, China*

[2] *National Space Science Center, Chinese Academy of Sciences, Beijing 100190, China*

## Summary

Due to ionosphere absorption and the interference by natural and artificial radio emissions, astronomical observation from the ground becomes very difficult at the wavelengths of decametre or longer, which we shall refer as the ultralong wavelengths. This unexplored part of electromagnetic spectrum has the potential of great discoveries, notably in the study of cosmic dark ages and dawn, but also in heliophysics and space weather, planets and exoplanets, cosmic ray and neutrinos, pulsar and interstellar medium (ISM), extragalactic radio sources, and so on. The difficulty of the ionosphere can be overcome by space observation, and the Moon can shield the radio frequency interferences (RFIs) from the Earth. A lunar orbit array can be a practical first step of opening up the ultralong wave band. Compared with a lunar surface observatory on the far side, the lunar orbit array is simpler and more economical, as it does not need to make the risky and expensive landing, can be easily powered with solar energy, and the data can be transmitted back to the Earth when it is on the near-side part of the orbit. Here I describe the Discovering Sky at the Longest wavelength (DSL) project, which will consist of a mother satellite and 6~9 daughter satellites, flying on the same circular orbit around the Moon, and forming a linear interferometer array. The data are collected by the mother satellite which computes the interferometric cross-correlations (visibilities) and transmits the data back to the Earth. The whole array can be deployed on the lunar orbit with a single rocket launch. The project is under intensive study in China.

## 1. Introduction

*Author for correspondence (xuelei@cosmology.bao.ac.cn).

Modern aastronomical observations are conducted over a very wide range of wavelengths, from radio, through infrared, optical, ultraviolet, X-ray, to gamma-ray, and further supplemented by non-electromagnetic observations such as neutrinos and gravitational wave events. Each new observational domain brought many unexpected discoveries and greatly changed our view of the Universe. However, at the long wavelength end (frequency below 30 MHz), the sky has barely been explored (see e.g. [1-3] for a summary of the early observational efforts), as the ionosphere severely distorts and absorbs the electromagnetic wave, and radio frequency interferences (RFI) are also very strong and nearly omnipresent, thanks to the reflection of the ionosphere.

A number of early space missions, such as the IMP-6 [4], Radio Astronomy Explore (RAE)-1 [5] and RAE-2 [6,7] made low frequency radio observations from space. Some solar system probes such as WIND, Cassini, etc. also carried low frequency radio payloads. The data collected by these satellites showed that the Earth have strong natural radio emissions at the kilometric wavelengths (note this radiation is at a longer wavelength than that of dark ages observation), and as the artificial sources are much stronger than the radio astronomical sources, even with the ionospheric absorption, man-made radio frequency interferences are visible from space.However, the angular resolutions of these single receiver satellites are very poor. It is time to consider a new space mission to explore and reveal the mysteries of this wave band.

There are a number of options for a low frequency radio space mission. Deploying satellites in orbit around the Earth is relatively simple and cheap to do, indeed the SunRise is such a mission [8]. but the RFI from the Earth would be a major problem. An array on the Sun-Earth L2 point allows all-time monitoring of the whole sky, but it is also constantly exposed to the radio emission from the Earth, though reduced in magnitude by the distance. Launching the satellites and maintaining the array configuration around the unstable L2 point, determining their relative positions in widely separated sky directions , and transmitting the data back all require substantial amount of resources both on board the satellite and on ground, making the mission quite hard. Radio observation from the far side lunar surface is another option. A first experiment has been carried out in 2018 by the Chang'e-4 (CE-4) lander [9]. For an array on the lunar surface, the imaging methods and tools developed for the ground-based radio astronomy could be readily used. However, a relay satellite is required to transmit the data back to the Earth, and furthermore, supplying energy for the lander during the half-month-long lunar night requires special power source such as radioisotope thermoelectric generator.

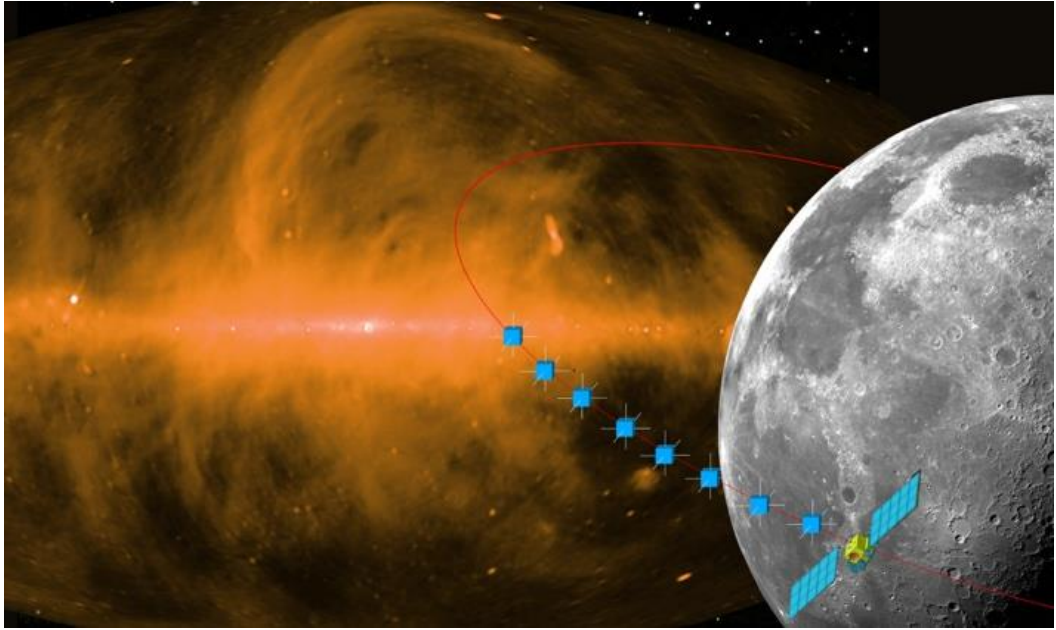

Figure 1. An artist's concept for the DSL mission.

Making radio observations from the orbit around the Moon has a number of advantages, which makes it easier to realize in near term. The part of orbit behind the Moon provides the perfect environment against the radio emissions from the Earth. The orbital period is about 2.3 hours for an orbit of 300 km altitude, so the energy can be easily supplied with solar power. The data can be transmitted back to the Earth when the satellite is on the near side, without the need of another relay satellite, and the complicated landing and deployment are also avoided. Even a single lunar satellite could measure the global average spectrum (i.e. the spectrum of the radiation averaged over the whole sky). The DARE/DAPPER mission is such a concept [10,11]. and may also detect strong radio sources by using the Moon as a moving screen. However, as the wavelength is very long, to achieve high angular resolution it is natural to consider an interferometer array [12,13,14]. There was an attempt to try the lunar orbit array during the CE-4 mission. Two micro-satellites (Longjiang-1 and-2) piggy-backed on the rocket which launched the relay satellite for the CE-4 mission [15], but unfortunately the Longjiang-1 satellite was lost shortly after the launch.

In this contribution, we describe the Discovering the Sky at Longest wavelength (DSL) mission [16,17,18], which employs a linear array of satellites on an inclined circular orbit around the Moon to make interferometry observations, and is currently under intensive study.

## 2. The Science Goals

As a first explorer mission, the main science objectives of the DSL are (1) to open up a new window of observation by mapping the sky and cataloguing the major sources at this wavelength, to reveal new astrophysical phenomena at this wavelength, and to discover the unknown unknowns; (2) to explore the dark ages and cosmic dawn by making high precision global spectrum measurements.

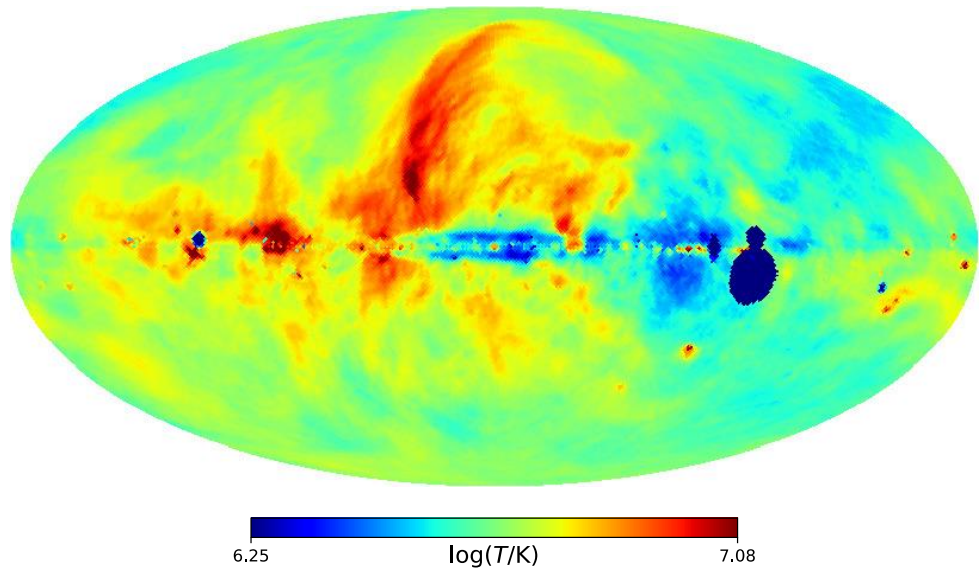

Figure 2. A simulated sky map at 3 MHz (from Y. Cong et al., *in preparation*)

With the high-resolution map, we will see various objects shining in this unexplored wave band. It could be stars and planets, nebulae, quasar jets and relics, galaxy cluster radio halos, to name just a few. The observation at this wavelength will provide new information about these objects. As a simple example of what can be learned from such observations, in Fig. 2, we show a simulated sky map at the 3 MHz, based on an extrapolation of maps at higher frequencies, with the emissivity given by an exponential disk model, and the expected absorption from the interstellar medium (ISM) using the NE2001 model [19]. The ISM is the matter that exists in the space between the star systems in the Galaxy, including gas in ionic, atomic and molecular form, as well as dust and cosmic rays. It provides the material from which the stars and planets form, and it is also constantly replenished with newly accreted gas as well as material and energy from stellar winds, planetary nebulae and supernovae. The ISM plays a crucial role in astrophysics, the very low frequency observations of DSL will provide an excellent means of tracing the diffuse ionized ISM. A conspicuous feature of the figure is that the galactic plane, which is the brightest part of the sky at higher radio frequencies, becomes the darkest part of the sky (shown as blue-black colour), thanks to the strong absorption which becomes significant at this low frequency. This allows us to see the ionized gas around us. The local bubble, in which our solar system resides, is especially of interest, as it may have some impact on the solar system for the last few million years. The dark spot in the map, which is created by a nearby HII region, may also provide a perfect place to study the production and propagation of the cosmic ray particles, as the radio emission of the cosmic ray electrons from behind the region are blocked.

The evolution of the Universe from the end of the hot Big Bang, through the dark ages, to the formation of the first luminous objects are only speculated. It remains a gap in our knowledge about the cosmic history. Furthermore, the large comoving volume and perturbations still undergoing linear evolution during these epochs are also precious trove of information for the high precision study on the primordial fluctuations and cosmic origin. However, the strong foreground at low frequency means that extremely high sensitivity is required in order to measure these primordial fluctuations, this is far beyond the capability of the present experiment. A first step toward the study of the dark age and cosmic dawn can be made with the global spectrum measurement. The signal-to-noise ratio of a global spectrum detection is independent of the receiver collecting area, hence could be carried out with a single antenna. During cosmic dawn the Lyman alpha photons produced by the first stars couple the spin temperature of the hydrogen gas to its kinetic temperature, and the latter is cooled by the expansion of the Universe and heated by X-ray radiation of first black holes. These processes may produce a deep absorption trough in the global signal (Fig.3)[20,21], and features in the global spectrum have been searched by a number of ground-based experiments, such as the EDGES, SCI-HI, SARAS and so on. Unexpectedly, an absorption as deep as twice that allowed by standard theory was reported by the EDGES experiment [22]. It has been speculated that such features could be explained by some exotic processes, such as baryon-dark matter scattering [23], but it is crucial to verify the result with an independent experiment which is least affected by various systematic effects, e.g. ground plane reflection [24]. In a lunar orbit experiment the problems of ionosphere distortion can be avoided, and although the Moon does reflect radio waves, the much larger height of the satellite means that it is unlikely to produce fake absorption features on the scales of interest. We note here that although the observation on lunar orbit has these advantages, the whole measurement is still highly challenging, because no antenna has a completely frequency-independent beam, and the spatial structure may be convoluted to the measured global spectrum to produce false absorption features. Sophisticated data analysis method needs to be developed to reduce such effect and check the validity of the detected features [25]).

Maps made by the DSL at lower frequencies would be useful for further studies on the dark age, revealing for the first time the radiation sources at these frequency ranges, and help with design of future larger missions.

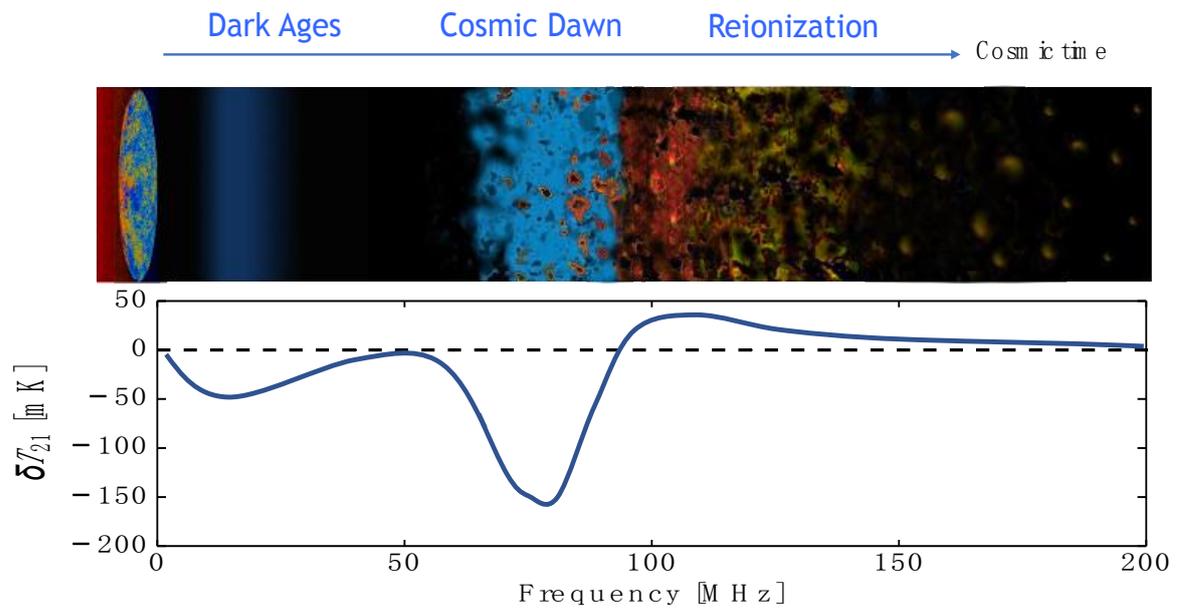

Figure 3. The 21 cm global spectrum (*lower panel*) and fluctuation signals (*upper panel*) throughout the cosmic history.

Finally, the mission may detect varying radiation from within the solar system, such as the Type II/III solar bursts, or the low frequency radio burst of the Jupiter and other planets, though the array is perhaps too small to detect the burst from other stars or exoplanets. For a linear array such as the DSL, the spatial structure of such varying source could be obtained only in one direction. Still, the angular resolution of the array may allow it to resolve heretofore unknown compact structure in the burst.

Throughout the history of radio astronomy, there are numerous examples of unexpected discoveries, such as the detections of radio emission from celestial radio sources in continuum (e.g., pulsars, Fast Radio Bursts) and spectral lines (e.g., OH masers). Each radio astronomy instrument built with at least one unique capability--sensitivity, spectral, time or angular resolution, spectrum coverage, etc., is practically guaranteed of outstanding discoveries. There is every reason to expect that a prospective space-based ultra-long wavelength facility will continue the trend of delivering unexpected discoveries.

# 3. Basic Concept of the Lunar Orbit Array

The proposed array consists of a constellation of satellites circling the Moon on nearly-identical orbits, forming a linear array while making interferometric observations. The linear configuration allows the array to stay in a relatively stable configuration in orbit, and relative positions of each satellite can be measured accurately with very limited instrumentation; a star sensor camera for angular measurement, and a microwave communication link for distance measurement, the latter is also used for inter-satellite data communication and synchronization. The present design of the array adopts a circular orbit of 300 km height which is shown by orbital simulation to be sufficiently stable against lunar gravity perturbations. A lower orbit has the advantage that it allows longer shielding time against the Earth, this is especially important below 0.5 MHz, as the AKR region extends several Earth radii[26], but due to the greater perturbation of the lunar gravity field, the orbit management is harder.

The constellation will include a mother satellite and a number of (tentatively set as 9) daughter satellites. One of the daughter satellites will be dedicated to the global spectrum measurement at the higher frequency band of 30 - 120 MHz, with the aim of probing the 21cm signal from the cosmic dawn. The other 8 daughter satellites will perform the spectrum and interferometric imaging observation below 30 MHz. The mother satellite, flying in the front or end of the linear array, will collect the digital signals from each daughter satellite and perform the interferometric cross-correlations, store the results, and transmit the data back to the Earth when it is in the near-side part of the orbit. The daughter satellites will be docked on the mother satellite during launch and lunar transfer, then sequentially released after entering the lunar orbit to form the linear array. The assembly of mother and daughter satellites is shown in Fig. 4. Each daughter satellite is equipped with its own propulsion system, so that it could make small adjustment of position in orbit.

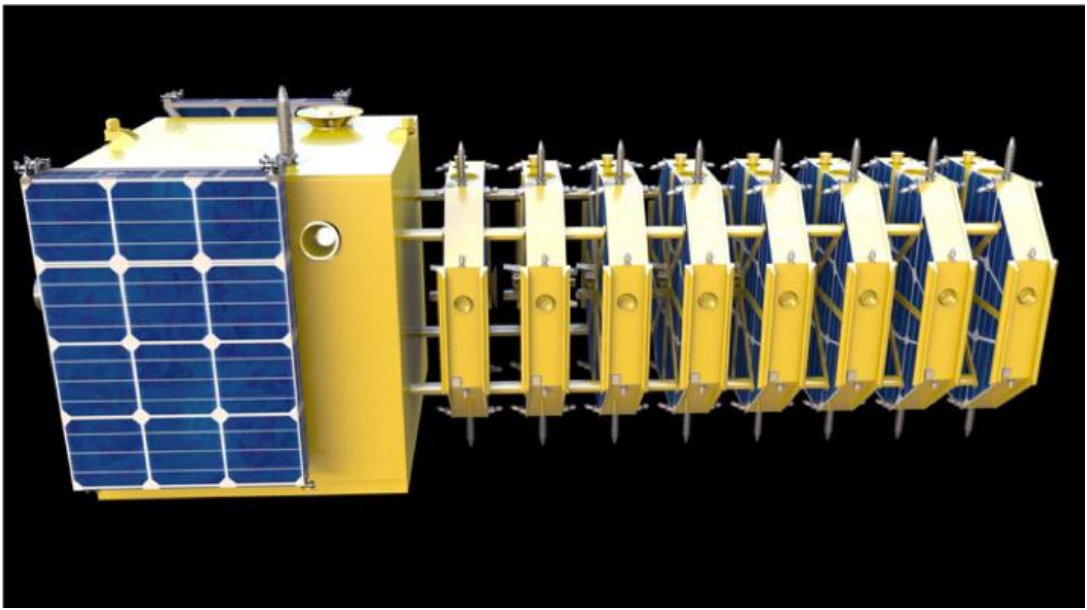

Figure 4. The mother and daughter satellites assembly. The total weight at launch is about 1400kg, with 600kg propellant. The dry weight of the mother satellite is about 330kg, and each daughter satellite 50kg.

The upper part of the mother satellite is equipped with deployable solar panels, a high gain antenna for ground communication, and other sensors and antennas. The lower part of the mother satellite carries the daughter satellites. After entering the lunar orbit and being released, the satellites will naturally have some velocity differences under the irregular gravity field of the Moon. However, the formation is automatically controlled to keep the linear array stable.

Each daughter satellite is a thin octagonal slab, with a diameter of about 1 meter, and the slab surface allows relatively large area for solar cells. The regular smooth geometry is designed to simplify the modeling of the antenna response. One of these is dedicated to the high precision measurement of the global spectrum in the high-band (30-120 MHz) for cosmic dawn. A monopole cone antenna with a ground plate is selected for the measurement of this frequency band. The antenna is designed to have a nearly frequency-independent beam and to be thermally stable and mechanically rigid. A precise calibration system will be embedded into the receiver as a core module, and differential measurement will be used in the receiving system. In order to detect the trough of 21 cm absorption, the goal is to reach an error of 5 mK on the brightness temperature at 75 MHz within the mission duration; this would allow a detection of the cosmic dawn signal at SNR > 10. Reaching this accuracy requires > 60 dB pass-band calibration. For foreground subtraction, the full band needs to be observed with a spectral resolution of 1 MHz.

Each of the other 8 daughter satellites carries three orthogonal short dipole antennas, a receiver, a digitizer, and a low frequency interferometer and spectrometer (LFIS). The daughter satellites are stabilized with respect to the Moon, with a pointing accuracy of the three axes better than 1 °, attitude stability better than 0.1 °/s, and attitude measurement better than 0.01 °. Limited by the inter-satellite communication bandwidth, a selection of 30 narrow bands (each with 8 kHz bandwidth) within the 0.1 - 30 MHz range will be used for interferometry.

The inter-satellite communication between each daughter satellite and the mother satellite is provided by a Ka-band microwave link, which also serves to synchronize the clocks on each satellite. In addition, it also measures the distance from each daughter satellite to the mother satellite. The angular position of a daughter satellite with respect to the mother is measured by having the star sensor camera on the daughter to take photographic shots in the direction of the mother, which carries an LED light array for identification against the star background. This system (called the inter-satellite dynamic baseline apparatus, ISDBA) is required to achieve a positioning precision of 1 m in each direction (1/10 wavelength at 30 MHz), which translates to a ranging precision of 1 m, with the corresponding clock synchronization precision of 3.3 nanosecond, and an angular precision of 10 urad at 100 km.

As the constellation of satellites orbits the Moon, multiple baseline vectors are formed between the 8 daughter satellites, and generate concentric rings in the *uv* plane (*u, v,* and *w* are Cartesian coordinates in units of wavelength, with w usually pointing toward the target direction). The gaps between the rings can be partially filled by varying the spacing between the satellites, or by bandwidth synthesis. It should be emphasized that the satellites do not need to have fixed relative positions, but as long as the positions can be determined accurately enough, the interferometry could work. The short dipole antennas are sensitive more or less in all directions, and the Moon only shields a small fraction of the sky, therefore, there is a mirror symmetry between the two sides of the orbital plane that cannot be broken using the data from a single orbit alone. However, the orbital plane precesses for 360 degrees in 1.29 years, so after a few orbits, a three-dimensional distribution of the baselines will be formed, and thus sources can in principle be determined by combining these measurements from different aperture planes. After a full-period precession, the baselines will fill a 3D "doughnut" shaped structure, and a sky image can be reconstructed by linear inversion from the 3D visibility dataset [17].

The angular resolution of an interferometer array is determined by the longest baseline. However, at low frequencies, the resolution is also limited by the broadening effect induced by scattering in the ISM as well as in the interplanetary medium (IPM) [27]. The maximum baseline for DSL is set to be 100 km, corresponding to a resolution of 0.17 degree at 1MHz, which is better than the limit set by the scattering. Except for the radio recombination lines and molecular lines detections, most of the known radiation mechanisms at low frequencies such as the synchrotron and free-free radiation produce smooth continuum radiation, for which high spectral resolution is not needed. Also, the foreground subtraction of the redshifted 21 cm line also only requires a moderate spectral resolution. However, fine spectral resolutions are still required in order to keep frequency coherence, because the baselines are moving as the satellite array circulating the Moon.

Detection sensitivity at long wavelengths is strongly influenced by the confusion limit, which limits the identification and measurement of individual sources, and occurs when there is more than one source in every 30 synthesized beams. Based on an extrapolation of the VLA 74 MHz and the Parkes 80/178 MHz source counts, we estimate that at 1 MHz, the confusion limit will be reached after observing with 8 antennas for a few weeks, but at higher frequency the mapping sensitivity will be primarily limited by integration time, not by confusion.

# 4. Key Technologies

## (a) Satellite formation fly in lunar orbit

To ensure safety operation as well as to maximize observation time, aircular orbit of 300km altitude (2.3 hour period) is chosen. We find that an inclination angle of about 30° will allow a good 3D distribution of the baseline and a relatively fast precession of the orbit plane, and a good portion of the orbital time would allow the Earth being shielded. Orbit control and management includes deployment of daughter satellites, linear array keeping, and linear array reconfiguration. The control strategy is based on the objective of balancing and minimizing the fuel consumption of each daughter satellite.

The orbit data of the mother satellite is obtained by ground-based ranging and VLBI, while the relative positions between the satellites are determined by the onboard ISDBA system. With positions relative to the mother satellite measured, the absolute position (i.e. with respect to the Earth-Moon system) of each daughter satellite is also determined.

The control scenarios for the linear array fleet is as follows:

- Deployment of daughter satellites. When the satellite combination manoeuvred to the 300 km operation orbit, the daughter satellites are ejected one by one from the mother satellites with predetermined time intervals. The release delta-V is about 0.04m/s. After about 100 orbital periods (about 10 days), all the daughter satellites are released, forming a linear array configuration.
- Linear array fleet keeping. The main objective of fleet keeping control is to maintain a stable configuration and prevent collision. Each satellite is kept to fly within its safety box, and maintain a stable attitude.
- Linear array fleet reconfiguration. The length of baselines can be varied by reconfiguring the formation, the main considerision in this is to conserve the propellant, and balance the consumption on each satellite. Here a sketch is made of the reconfigurations process:

    Step1: keeps S1 stationary, S2-S8 shrinking;

    Step2: keeps S8 stationary, S1-S7 expanding.

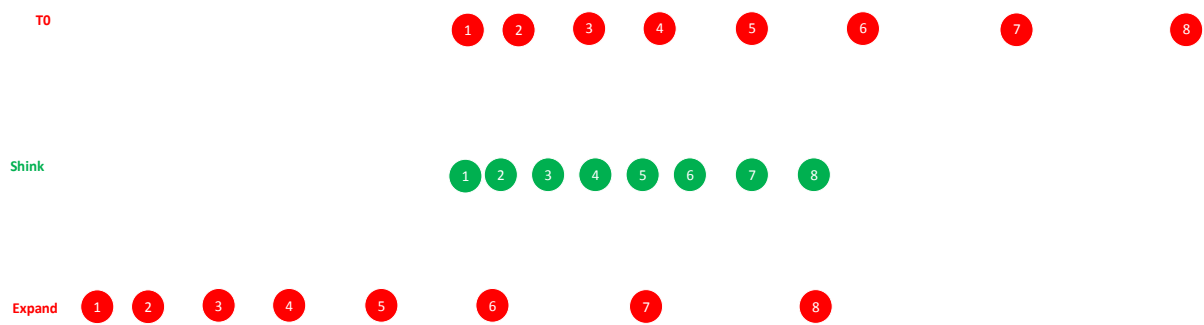

Figure 5. Formation reconfiguration strategy in response to the science requirements.

To ensure sufficient time for collision avoidance control, it is preferred to move the satellites one after another instead of having all of them move simultaneously. The control delta-V can be derived analytically through relative dynamics analysis. The linear array reconfiguration period is about 20days, each satellite needs 80 mm/s delta-V in each linear reconfiguration control period. For the whole mission there will be about 50 cycles of reconfigurations.

## (b) Precision Measurement of Relative Positions and synchronization

We have explored various technologies for the relative position determination and data communication. For example, laser technology can provide much larger inter-satellite communication bandwidth and higher precision for ranging and angular position measurement [28]，but it is more suitable for point-to-point communication, communicate with all daughter satellites simultaneously is much harder, and the mass, volume and power consumption requirements are also too high for our daughter satellites. At present, it appears that the microwave technology provides the best solution for our need.

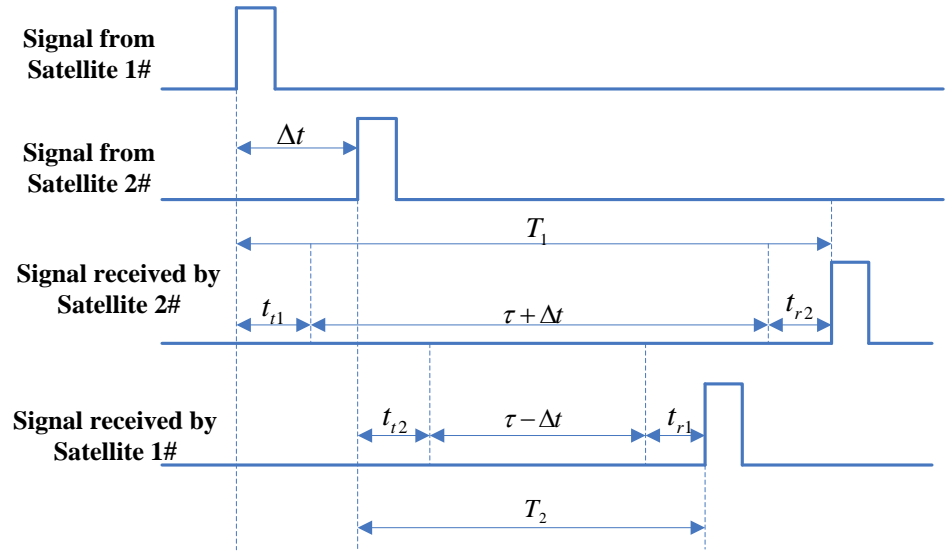

Figure 6. The principle of dual one way ranging (DOWR) measurement.

The inter-satellite communication as well as ranging and clock synchronization can all be realized with one microwave link by using the dual one-way ranging (DOWR) principle, and simultaneous operation with all satellites can be achieved easily with frequency division multiplexing. A similar system was developed for the Longjiang satellite experiment [15,29]. Ground test show that centimetre-sized precision can be achieved in ranging, and the precision of clock synchronization is less than 2 ns. DOWR [30,31] is a simple and effective way to measure the time difference, as shown in Fig. 6. Time difference multiplied by light speed is equal to the distance between two satellites. The communication frame, bit and phase can be recovered so that the local time difference is measured at the daughter side. Subsequently, the remote time difference measured by the mother is communicated through the inter-satellite link. The real time difference can then be computed from the remote and local time differences. Random jitter in the measurement is reduced by employing a Kalman filter. The distance between the mother and daughter is then obtained by multiplying the real time differences with speed of light. After this measurement, the clock on the daughter satellite is synchronized to the clock on the mother satellite by time discipline: a high-precision digital-to-analog converter (DAC) is employed to adjust the VCOCXO (Voltage Controlled Oven Controlled Crystal Oscillator) by a PID (proportion, integral, derivative) controller, as shown in Fig. 7. The ranging and clock synchronization are thus realized simultaneously.

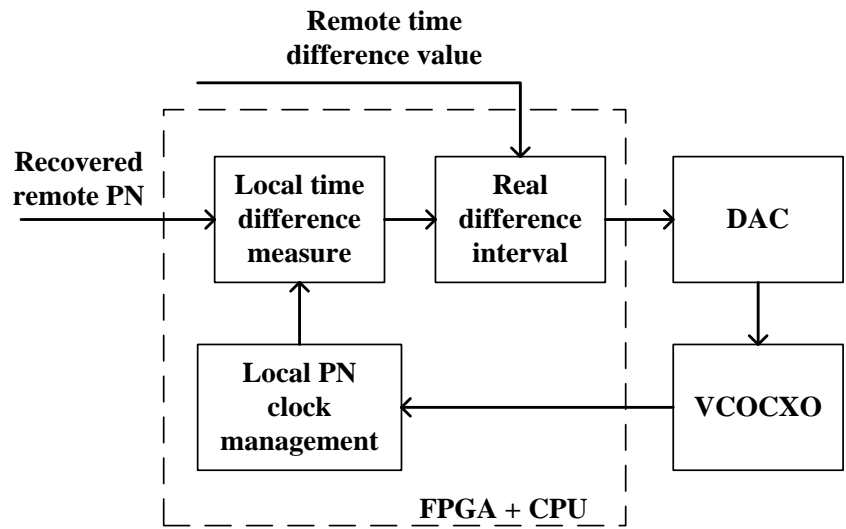

Figure 7. The time discipline process.

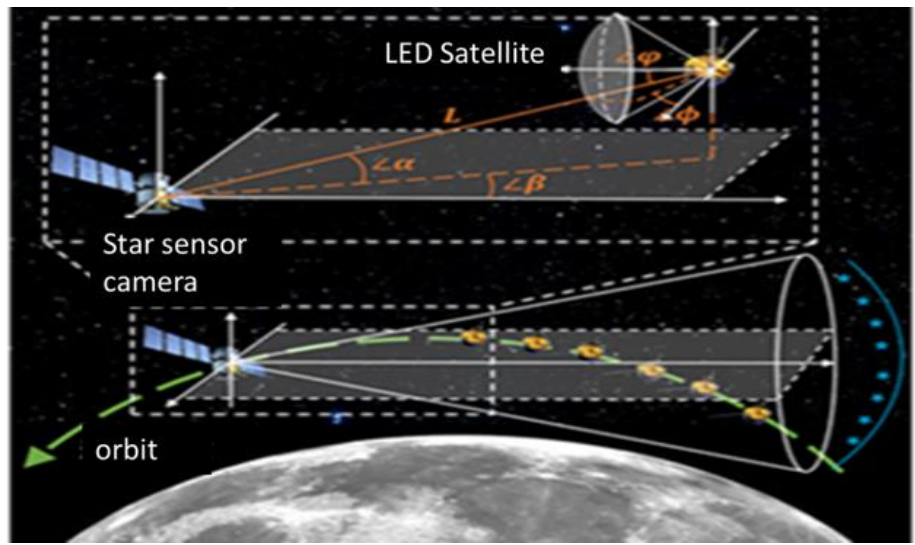

Figure 8. Angular measurement with star sensor camera and LED lights.

    The direction of the vector between the daughter and mother satellite is determined by locating the mother satellite position against the star background, and an LED light array (60 W for the 100 km baseline) is carried by the mother satellite as target, and each daughter satellite uses its star sensor camera to take the photograph (Fig. 8). This method of measurement is not affected by the mechanical installation error of the equipment and satellite attitude error as long as they are within the range to allow normal operation. The angular accuracy is mainly determinant by the angular precision of star sensor camera [32].

## (c) High precision calibration

Calibration of the radio receiver for the DSL mission faces two challenges: first, usually the radio telescope is first calibrated with a strong point source which can dominate the signal received by the antenna. But the sky at this band is very bright, we do not know what sources are there, so it is unclear whether there are suitable sources to serve as calibrator. Even if there are some very bright point sources, the electrically short antenna has very broad beam so it is not easy to isolate the effect of the calibrator source from the rest of the sky. Second, the extremely high precision required for the cosmic dawn and dark age measurement is itself very challenging to achieve. To cope with these problems, we consider using artificial calibration mechanism to ensure the success of the calibration.

To calibrate the instrumental phases of the distributed interferometer array elements, a digitally generated signal can be broadcasted from the mother satellite. After this relative phase calibration step, there is still an overall phase undetermined, which must be calibrated with an external source. However, with the relative phase calibration done, even if the astronomical source is dominating, it may still be possible to extract its fringes in the visibility data and determine the remaining phase.

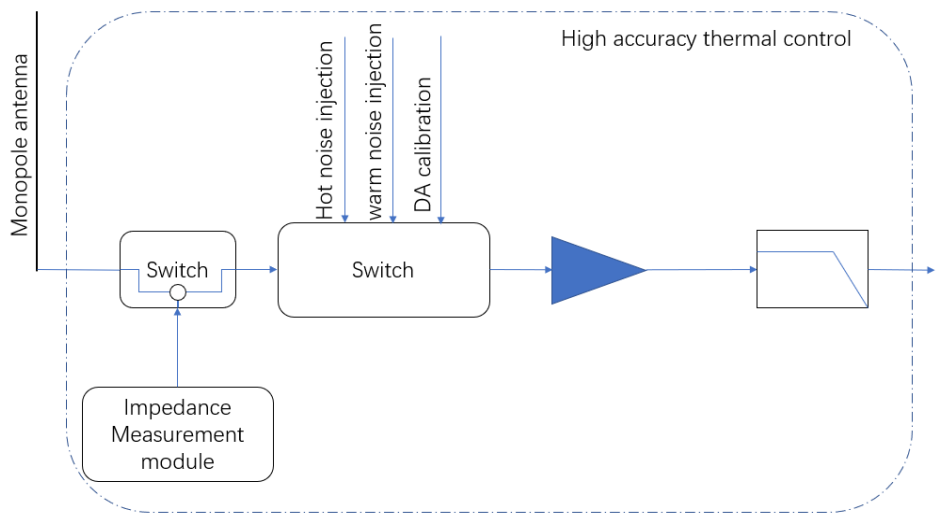

Figure 9. The receiver channel with calibration subsystem.

For the cosmic dawn spectrum measurement, the main concern in the design of antenna is to minimize the chromaticity, i.e. the variation of beam with frequency. The resonance frequency is outside the observation band for electrically short antenna, making the frequency response smoother. The receiver can be calibrated by switching to several different modes. For example, two-point calibration is performed by injecting hot and warm noise generated from highly stable white noise sources, to determine receiver amplitude response. The impedance mismatch within the system must also be considered: the antenna output and receiver impedances can be measured in-orbit to establish the receiver transfer function. Short/long cables with open, short and load terminators are also used to calibrate the system parameter. System stability is also improved by having a high precision thermal control subsystem which should provide good control of ±0.1℃. Figure 9 shows a diagram of the receiver channel with calibration subsystem. However, achieving the required precision is still a great challenge, which requires further study on the possible systematics and their mitigation.

## (d) Imaging algorithm with large field of view, 3D baseline distribution, and time-dependent blockage

The imaging for a lunar orbit array is facing a number of complications. The 8 daughter satellites will be each equipped with three, orthogonal dipole antennas, which are sensitive to the full, $4\pi$ steradians of the sky. If a telescope has a small FoV, the sky can be approximated as a plane and images synthesis is simply a Fourier inverse transform of the calibrated visibilities. Many methods have been developed to deal with large FoV [33], but these algorithms were all designed for ground-based arrays, for which there is always a ground-screen (either part of the instrument or the Earth itself) that restricts the FoV to at most half of the sky. Here sources on both sides of the orbital plane are in view, resulting in a 'mirror symmetry' problem. However, as the orbital plane processes in the gravitational field of the Moon and the Earth, a 3D distribution of baselines will be formed. This will complicate the sky reconstruction process as the so called "w-term" can not be neglected, but the mirror symmetry is broken. New imaging algorithm needs to be developed[17], which are likely to be computationally expensive as they will need to make use of the full 3-D distribution of $u,v,w$ points.

Another complication comes from the continuous motion of the space-based array. With an orbital period of 2.3 hours, the $u,v,w$ coordinates are changing much faster than a ground-based array. Hence, for the interferometric observation, the integration time must be very short (38 ms) to avoid time decorrelation. Also, unlike a ground-based array where the relative positions are stable and can be measured with very high precision, the baselines of our array are constantly changing and measured with limited precision, this will also degrade the imaging quality.

## (e) Electromagnetic interference (EMI) suppression and removal

With the RFIs from the Earth shielded by the Moon, the spacecraft itself becomes the main source of RFI. Self-generated RFI is best tackled in the design phase, as reducing these emissions in existing designs is very difficult and costly. The DSL working frequency range is 0.1 MHz – 120 MHz (interferometry in 0.1 MHz – 30 MHz and spectral in 30 MHz – 120 MHz), while the RFIs from the electronics also often fall within this low frequency band.

Strong RFIs are found at the switch frequencies of the DC-DC module, the frequency of the clock, data bus, etc. Systematic EMI attenuation techniques are applied to control the EMI from the sources level. In the design of the PCB boards, EMI simulation will be used to keep the signal integrity, EMI Filters will be used in the connectors to minimize the radiation from the cable connections, and the satellite platform and payload will have carefully designed EMI shielding structure. The RFI frequency management, e.g. by derive the various oscillators in different components from the same master oscillator, can reduce the RFI contamination to a limited number of frequencies. In addition, choosing narrow scientific observational frequency channels (down to kHz level, if the system budget allows this) is advantageous as this would allow excising narrow-band RFI.

In terms of algorithmic RFI mitigation approaches, there are many ways to detect and suppress RFI. Intermittent RFI can be detected by using a power detector or higher order statistics (kurtosis, cyclo-stationarity), although the later may require significant computing resources, and may not be available on-board. For multiple antennas on a spacecraft (two or three orthogonal dipoles polarization, or a number on monopole antennas) one can also use spatial filtering to suppress RFI. It may be advantageous to add an additional (small) reference antenna. As it is possible to find a direction of arrival (DoA) with just one set of co-located antennas (using goniopolarimetry), it is also possible to suppress signals from a certain DoA. This principle, to some extent, is also applicable to RFI generated by the spacecraft itself.

## 5. Conclusion

The ultra-long wavelength radio signals have great potentials for scientific breakthrough, especially for the study of the cosmic dawn and dark ages. Imaging the neutral hydrogen distribution during the dark ages can provide valuable information about the primordial density perturbations and the inflationary origin of the Universe, but this requires extremely large arrays on the far side of the Moon. A useful and practical first-step is to map the foreground which is necessary for the design of future dark ages experiments, and to measure the global spectrum with high precision, which gives a first peek of the cosmic dawn and dark age. This first step can be made in the coming decade with a lunar orbit array such as the DSL. The technologies required by such a mission are being developed. International collaboration would enrich and advance these researches.
*Phil. Trans. R. Soc. A.*

The coming decade offers a unique opportunity for RF-quiet astronomical observations from the lunar far side, before further development of lunar assets compromises the RF-quiet character in the long-term.

**Authors' Contributions**

XC is the PI of the project and the primary author of this paper. JY is the technology chief and contributed to the discussion on the key technologies. LD, FW, LW and LZ contributed to the part on key technologies, YX contributed to the part on science and the part on key technologies. All authors read and approved the manuscript

**Competing Interests:** The authors declare that they have no competing interests.

**Funding Statement:**
The DSL project is supported by the Chinese Academy of Science Strategic Priority Research Program XDA15020200. Additional support for the research at NAOC is provided by Chinese Academy of Science through grant QYZDJ-SSW-SLH017, NSFC through grants 11633004, 11973047, 11761141012, 11653003.

**Acknowledgments**
We thank Linjie Chen, Ailan Lan, Maohai Huang, Qizhi Huang, Yuan Shi, Guanqun Song, Shijie Sun, Zhugang Wang, Ji Wu, Bin Yue, Mo Zhang, Fei Zhao, Chengguang Zhu for discussions.